\documentclass[10pt,letterpaper]{article}

\usepackage{moreverb}
\usepackage{amsmath,amssymb}

\usepackage[
colorlinks,bookmarksopen,bookmarksnumbered,citecolor=red,urlcolor=red]{hyperref}

\usepackage{enumerate}
\usepackage[english]{babel}
\usepackage{xr}
\usepackage{amsfonts}
\usepackage{booktabs}
\usepackage{diagbox}
\usepackage{longtable}
\usepackage{epstopdf}
\usepackage{bbm}
\usepackage[shortlabels]{enumitem}
\usepackage{setspace}
\usepackage{lscape}
\usepackage{listings}
\usepackage[utf8]{inputenc}
\usepackage[english]{babel}
\usepackage{units}
\usepackage{algorithm}
\usepackage{algpseudocode}
\usepackage{pdfpages}
\usepackage{caption}
\usepackage{mathtools}
\usepackage[margin=0.6in]{geometry}

\newtheorem{theorem}{Theorem}[section]

\begin{document}
	
	\begin{center} 
		{\Large
		
			\centering\textbf{\newline{Multivariate Sparse Group Lasso Joint Model for Radiogenomics Data}} 
		}
		\newline
		\\
		Tiantian Zeng \textsuperscript{1*}, 
		Md Selim \textsuperscript{2},
		Jie Zhang \textsuperscript{3},
		Arnold Stromberg \textsuperscript{1},
		Jin Chen\textsuperscript{2,4}, 
		Chi Wang \textsuperscript{1,4*}
	\end{center}
		\textbf{1} Department of Statistics, University of Kentucky, Lexington, KY, US.
		\\
		\textbf{2} Department of Computer Science, University of Kentucky, Lexington, KY, US.
		\\
		\textbf{3} Department of Radiology, University of Kentucky, Lexington, KY, US.
		\\
		\textbf{4} Department of Internal Medicine, University of Kentucky, Lexington, KY, US.
		\\
		
		*Correspondence to: t.zeng@uky.edu; chi.wang@uky.edu.

	
	\section*{Abstract}
Radiogenomics is an emerging field in cancer research that combines medical imaging data with genomic data to predict patients' clinical outcomes. In this paper, we propose a multivariate sparse group lasso joint model to integrate imaging and genomic data for building prediction models. Specifically, we jointly consider two  models, one regresses imaging features on genomic features, and the other regresses patient’s clinical outcome on genomic features. The regularization penalties through sparse group lasso allows incorporation of intrinsic group information, e.g. biological pathway and imaging category, to select both important intrinsic groups and important features within a group. To integrate information from the two models, in each model, we introduce a weight in the penalty term of each individual genomic feature, where the weight is inversely correlated with the model coefficient of that feature in the other model. This weight allows a feature to have a higher chance of selection by one model if it is selected by the other model. Our model is applicable to both continuous and time-to-event outcomes. It also allows the use of two separate datasets to fit the two models, addressing a practical challenge that many genomic datasets do not have imaging data available.  Simulations and real data analyses demonstrate that our method outperforms existing methods in the literature.


\section{Introduction}
In recent years, among the increasing number of cancer studies, radiogenomics had emerged as a new research direction. For a cancer patient, ordinarily, some imaging features such as tumor sizes and locations will be addressed from patients' CT images, MRI images, etc. These imaging features could be used for diagnosis and treatment, which forms a significant part of cancer clinical protocols \cite{fass2008imaging}. Besides, recently a lot of research work has shown that genomic data should be applied for cancer diagnosis and treatment prediction \cite{davies2002mutations, sotiriou2009gene, balmain2003genetics}.  However, genomic mechanism is complicated, and genomic data has too many variables compared to the limited sample size. Thus the redundant information could lead to limited prediction power \cite{george2015comprehensive}. Therefore, the new research direction, radiogenomics, which integrates radiomics, genomics and clinical data, is a promising research field that allows researchers to derive predictive association maps, and could assist better in cancer diagnosis and treatment prediction \cite{lam2018radiogenomics}. Specifically, combining the imaging and genomics data is a better way to assist predicting the clinical outcome, since the dataset is usually not of enough sample size to do feature selection, considering the large number of genomic and imaging features. 

The challenge in radiogenomics data analysis is not only the limited sample size, but also the lack of large-scale datasets with all three types of data, i.e. imaging, genomic, and clinical data, available  \cite{bai2016imaging, mazurowski2015radiogenomics}. There are many public datasets only has genomic information but not imaging information. The largest public dataset for radiogenomics is published in Bakr \textsl{et al.} (2018) \cite{bakr2018radiogenomic}, with only around one-hundred patients in total. In addition, there are available genomic data and clinical data of patients with non-small cell lung cancer(NSCLC) in TCGA research network, and the sample size is over five-hundred. However, imaging data is not available in TCGA. Hence, it is desired to develop a model that can not only integrate genomics information and imaging information, but also can enable utilizing the partially available data. 

For the single type of data analysis, e.g. using genomic data to predict clinical outcomes, Lasso model and group lasso model are widely applied, given the fact that genomic data is high-dimensional, i.e., the number of predictors is much larger than the number of observations. The Lasso is a popular method to solve the problem of high-dimensional predictors with sparsity \cite{tibshirani1996regression}. It applies regularization penalties to do the shrinkage of the coefficients of irrelevant predictors. However, regular Lasso does not perform well for the correlated features \cite{hastie2015statistical}, which is very common in genomics data since some genes from the same biological pathway could have high correlation with each other. Yuan and Lin \cite{yuan2006model} proposed the group lasso method by introducing an additional penalty term for each predictor group. Then in order to select both group and within-group predictors, Simon et al. proposed a 'sparse-group lasso' model, which is a convex combination of the lasso and group-lasso penalties \cite{simon2013sparse}, and it can also be extended for the survival outcome. However, this model only applies to the univariate response, and the other limitation is that it does not work for overlapped grouping. Then, Li et al. proposed a multivariate sparse group lasso model for the case of multiple responses \cite{li2015multivariate}, but the limitation of the model is that it only works for linear association.

Therefore, with the inspiration from the literature, we propose to build a joint model, which is an integration of two models, that is, the models between imaging and gene expression, and between clinical outcome and gene expression correspondingly. Based on the sparse group lasso panelties that allows variable selection at both intrinsic group level and individual feature level, we further introduce a weight  to each panelty term of an individual feature  to enable conveying the information of feature selection from one model to the other. Therefore, our model allows leveraging the information on the association between imaging and genomic features to better select genomic features for the prediction of clinical outcomes. We implement an accelarated generalized coordinate descent algorithm to efficiently perform feature selection and model fitting. The performance of our method is evaluated by simulation studies and real data analysis.

\section{Method}
\subsection{Joint Model for Continuous Outcome}
The data structure of radiogenomics study consists of three parts, imaging, genomics, where we use gene expression, and clinical outcome. Assuming linear relationships for model derivation at first, we construct two models, one between imaging data (Y) and gene expression data (X) , and the other model is between clinical outcome (Z) and gene expression (X). It is assumed that the imaging data matrix Y is a matrix with n samples and q variables, the covariate matrix X is with n samples and p variables, and the clinical outcome response is a one-dimensional vector. The relationships are outlined below. 
\begin{center}
	Model 1, Imaging (Y) vs. Gene expression (X)  \\
\end{center}
\begin{equation}
	Y = XB+W
\end{equation}
\begin{center}
	Model 2, Clinical outcome (Z) vs. Gene expression (X)  \\
\end{center}
\begin{equation}
	Z = XG+W'
\end{equation}
where $Y=(y_1, ..., y_q) \in R^{n \times q}$, $Z \in R^{n}$, $X = (x_1, ..., x_p) \in R^{n \times p}$, and the coefficient matrix $B=(\beta_{jk})_{p \times q} \in R^{p \times q}$, $G=(\gamma_j)_{p \times 1} \in R^{p}$. \\
$W = (w_1, ..., w_q) \in R^{n \times q}$ is the matrix of error terms for Model 1, with each term $w_k \sim N(0, \sigma_k^2 I_{n \times n})$, k =1, ..., q, and $W'  \in R^n$ is the vector of error terms for Model 2, with $w' \sim N(0, \sigma'^2 I_{n \times n})$. 

This joint model enables integration of radiomics data, genomics data and clinical data.

We also include grouping in our model to facilitate the feature selection procedure and thus improve the model prediction. Gene expressions can be grouped by biological pathways, and imaging features can be grouped by categories, such as tumor size, shape morphology, enhancement texture, etc. Both groupings are pre-defined. Define the group structure $\textit{g} = \{B_1, ..., B_G \}$.

For Model 1, the penalized optimization problem is defined as
\begin{equation}
	L_1(B) = \arg \min_B \frac{1}{2n} {\parallel Y-XB \parallel}_{2}^{2} + \sum_{1 \leq j \leq p, 1 \leq k \leq q}\lambda_1{|\gamma_j^{*}| }^{\alpha} |\beta_{jk}| + 
	\sum_{g \in \textit{g}} \lambda_{1g}{\parallel \gamma_g^* \parallel_2^{\alpha} } {\parallel B_{g} \parallel}_2
\end{equation}
For Model 2, the penalized optimization problem is defined as
\begin{equation}
	L_2(G) = \arg \min_G \frac{1}{2n} \parallel Z-XG \parallel^2_2 + \sum_{1\leq j\leq p} \lambda_2 {|{\beta}_{j.}^*|}^{\alpha} |\gamma_j| + \sum_{g \in G} \lambda_{2g} {\parallel{\beta}_g^*\parallel}_2^{\alpha} \parallel G_g \parallel_2
\end{equation}
where for Model 1, $\gamma_j^{*} = -\tilde{\gamma}_{j} / (\frac{1}{P}\sum_{j=1}^P (-\tilde{\gamma}_{j}))$, while $\tilde{\gamma}_{j} = \log_{10} |\gamma_{j}|$ and $\gamma_j$ is the regression coefficient obtained from the Model 2, then define $\tilde{\gamma}_{j}=-0.01$ if $\tilde{\gamma}_{j} \geq 0$, and $\tilde{\gamma}_{j} = -2$ if $\tilde{\gamma}_{j}<-2$. Also, ${\parallel \tilde{\gamma_g}  \parallel}_2 = \sqrt{\sum_{\gamma_{j}  \in G_g} \tilde{\gamma_j}^{2}}$, and $\parallel \gamma_g^* \parallel_2 = {\parallel \tilde{\gamma_g}  \parallel}_2 / (\frac{1}{G}\sum_{g=1}^G {\parallel \tilde{\gamma_g}  \parallel}_2)$. 

For Model 2, $\beta_{j.}^* = -\tilde{\beta}_{j \cdot} / (\frac{1}{P}\sum_{j=1}^P (-\tilde{\beta}_{j \cdot}))$, while $\tilde{\beta}_{j \cdot} = \log_{10}  (\max \limits _ {1 \leq k \leq q} |\beta_{jk}|)$, and $\beta_{jk}$ is the regression coefficient obtaind from the Model 1, then define $\tilde{\beta}_{j \cdot}=-0.01$ if $\tilde{\beta}_{j \cdot}>=0$, and $\tilde{\beta}_{j \cdot} = -2$ if $\tilde{\beta}_{j \cdot}<-2$. Also, $\parallel \tilde{\beta}_g \parallel_2 = \sqrt{\sum_{\tilde{\beta_{j \cdot}}  \in B_g}  \tilde{\beta_{j \cdot}}^2}$, and $\parallel \beta_g^* \parallel_2 =  {\parallel \tilde{\beta_g}  \parallel}_2 / (\frac{1}{G}\sum_{g=1}^G {\parallel \tilde{\beta_g}  \parallel}_2)$. 

$\alpha$ is a parameter that determines joint model or separate model, $\alpha \ge 0$. 

In the defined joint model, the key point is to exploit the parameters obtained from one model, that are, $\gamma_j^{*}$ and $\beta_{j.}^*$ in the penalty term for each individual feature and also the grouping feature to the other model. According to the definition, $\gamma_j^{*}$ is a normalized regression coefficient from Model 2. With this penalty term, the model will be able to select features with information conveyed from Model 2. The larger the regression coefficient value from Model 2, the smaller $\gamma_j^{*}$ and thus the penalty term $\lambda_1 \gamma_j^{*}$ will be. Therefore, this feature will be selected with a higher possibility. Similarly, $\beta_{j.}^*$ is the normalized regression coefficient from Model 1. The larger the regression coefficient value from Model 1, the smaller this normalized regression coefficient and its penalty term $\lambda_1 \beta_{j.}^*$ will be. As a result, this feature will be selected with a higher possibility as well. 

The reason for taking logarithm of the regression coefficient and adding the range limit when doing normalization is to narrow the deviations among different obtained coefficient values, which is important since significant deviations will considerably alter the penalty term and subsequently the feature selection. 

In summary, the joint model helps us to integrate the three parts of the data structure and build the model iteratively. It allows us to (i) use one model to enhance the feature selection procedure of the other model, (ii) locate the corresponding genomic features that are associated with imaging features, and (iii) anchor the corresponding genomic features associated with the clinical outcome. Benefiting from these advantages, the joint model, when implemented into practical scenarios, is believed to achieve better prediction of the clinical outcome. 

The solution to our joint model is shown below. The solution and proof we used for reference are provided in Ref. \cite{li2015multivariate}. The detailed proof is shown in the Supplementary Materials.
\begin{theorem}
	For an arbitrary group structure \textit{g} on B, let $\hat{B}$ be the solution to Model 1, and $\hat{\beta_{jk}}$ be its $jk$-th element, 
	\begin{enumerate}
		\item
		\begin{equation}
			\hat{\beta_{jk}} = \frac{sgn(S_{jk})(|S_{jk}|-n\lambda_{1}{|\gamma_j^{*}| }^{\alpha})_{+}}{\parallel x_j \parallel_2^2 + n\sum \limits_{g \in \textit{g}: \beta_{jk} \in B_g} (\lambda_{1g} \parallel \gamma_g ^{\ast} \parallel_2^{\alpha} ) / \parallel \hat{B_g} \parallel_2}
		\end{equation}
		if all the groups $B_g \in \textit{g}$ containing $\beta_{jk}$ satisfy
		\begin{equation*}
			\sqrt{\sum\limits_{{jk:\beta_{jk} \in B_{g0}}} (|S_{jk}|/n - \lambda_{1}|\gamma_j ^{\ast}|^{\alpha} )_{+}^2 } > \lambda_{g0}
		\end{equation*}  
		\item $\hat{\beta_{jk}}=0$ if any of the groups $B_g \in \textit{g}$ containing $\beta_{jk}$ satisfies
		\begin{equation}
			\sqrt{\sum\limits_{{jk:\beta_{jk} \in B_{g0}}} (|S_{jk}|/n - \lambda_{1}|\gamma_j ^{\ast}|^{\alpha} )_{+}^2 } \leq \lambda_{g0}
		\end{equation}  
		where $S_{jk}=x_j^T(Y-X\hat{B}_{(-j)})_{\cdot k}$, with $\hat{B}_{(-j)}$ being j-th row of $\hat{B}$ replaced by zeros, the $\cdot k$ subscript refers to the k-th column of a matrix;
	\end{enumerate}
	
	Similarly,  for an arbitrary group structure $\textit{g}$ on G, let $\hat{G}$ be the solution to Model 2, and $\hat{\gamma_{j}}$ be its $j$-th element,
	\begin{enumerate}
		\item 
		\begin{equation}
			\hat{\gamma_{j}} = \frac{sgn(S_{j}')(|S_{j}'|-n\lambda_{2}|\beta_{j.} ^{\ast}|^{\alpha})_{+}}{\parallel x_j \parallel_2^2 + n\sum\limits_{g \in \textit{g}: \gamma_{j} \in G_g} (\lambda_{2g} \parallel \beta_g ^{\ast} \parallel_2^{\alpha})/ \parallel \hat{G_g} \parallel_2}
		\end{equation}
		if all the groups $G_g \in \textit{g}$ containing $\gamma_j$ satisfy
		\begin{equation*}
			\sqrt{\sum\limits_{j:\gamma_{j} \in G_{g0}} (|S_{j}'|/n - \lambda_{2}|\beta_{j.}^ {\ast}|^{\alpha} )_{+}^2 } > \lambda_{g0}'
		\end{equation*}
		\item $\hat{\gamma_{j}}=0$ if any of the groups $G_g \in \textit{g}$ containing $\gamma_j$ satisfies 
		\begin{equation}
			\sqrt{\sum\limits_{j:\gamma_{j} \in G_{g0}} (|S_{j}'|/n - \lambda_{2}|\beta_{j.}^ {\ast}|^{\alpha} )_{+}^2 } \leq \lambda_{g0}'
		\end{equation}
		where $S_{j}'=x_j^T(Z-X\hat{G}_{(-j)})$, and $B_{g0}$ and $\lambda_{g0}$ means some group $B_{g0} \in \textit{g}$ with a tuning parameter $\lambda_{g0}$.
	\end{enumerate}
	
\end{theorem}

Since these are not closed form solutions and numerically solving this solution iteratively is very time consuming, thus we switch to use the accelarated generalized coordinate descent algorithm, derived from the algorithm presented in Ref. \cite{simon2013sparse}.

Define $r_k^{(-j)}$ as 
\begin{equation}
	r_k^{(-j)} = Y_{k} - \sum_{j' \neq j} X_{j'}\beta_{j'k}
\end{equation}
where $Y_k$ refers to the k-th column of Y matrix, $X_j'$ refers to the j'-th column of X matrix, and $\beta_{j'k}$ refers to the j'-th row and k-th column of the coefficient matrix. 

Then the unpenalized loss function can be written as 
\begin{equation}
	l(r_k^{(-j)}, \beta_{jk}) =  \frac{1}{2n}\sum_{k=1}^{q} \parallel 	r_k^{(-j)}-X_{j}\beta_{jk} \parallel_2^2
\end{equation}

Referring to the commonly used modern method of gradient descent \cite{simon2013sparse, nesterov2003introductory, nesterov2013gradient}, we can majorize our loss function in Eqn(10) by
\begin{equation}
	l(r_k^{(-j)},  \beta_{jk}) \leq l(r_k^{(-j)},  \beta_{jk}^0) + (\beta_{jk}-\beta_{jk}^0)^T \bigtriangledown l(r_k^{(-j)}, \beta_{jk}^0)  + \frac{1}{2t}\parallel \beta_{jk}-\beta_{jk}^0 \parallel^2_2
\end{equation}
where $ \beta_{jk}^0$ is updated from the last step, and taken as a centered point here, and t is sufficiently small such that the quadratic term dominates the Hessian of our loss function  \cite{simon2013sparse}. 

Adding our penalty to the loss function, our penalized optimization problem becomes
\begin{multline}
	L_1(B)' = l(r_k^{(-j)},  \beta_{jk}^0) + (\beta_{jk}-\beta_{jk}^0)^T \bigtriangledown l(r_k^{(-j)}, \beta_{jk}^0)  + \frac{1}{2t}\parallel \beta_{jk}-\beta_{jk}^0 \parallel^2_2 \\
	+ \sum_{1 \leq j \leq p, 1 \leq k \leq q}\lambda_1{|\gamma_j^{*}| }^{\alpha} |\beta_{jk}| + 
	\sum_{g \in \textit{g}} \lambda_{1g}{\parallel \gamma_g^* \parallel_2^{\alpha} } {\parallel B_{g} \parallel}_2
\end{multline}

Since $ \beta_{jk}^0$ is updated from the last step, we take $ l(r_k^{(-j)}, \beta_{jk}^0)$ as a constant in terms of optimization, thus minimizing $L_1(B)'$ is equivalent to minimizing 
\begin{multline}
	\tilde{L_1(B)} = \frac{1}{2t} \parallel \beta_{jk}- (\beta_{jk}^0 - t \cdot \bigtriangledown l(r_k^{(-j)},\beta_{jk}^0)) \parallel_2^2 + \sum_{1 \leq j \leq p, 1 \leq k \leq q}\lambda_1{|\gamma_j^{*}| }^{\alpha} |\beta_{jk}| \\ + 
	\sum_{g \in \textit{g}} \lambda_{1g}{\parallel \gamma_g^* \parallel_2^{\alpha} } {\parallel B_{g} \parallel}_2
\end{multline}

For $\tilde{L_1(B)}$, combining the subgradient condition with basic algebra, and as \cite{gordon2012karush} has shown the satisfying Karush-Kuhn-Tucker conditions for Lasso and group Lasso, we get that the group containing $\beta_{jk}$ will be zero, i.e.$\hat{B_{g0}}=0$ if the following inequality is achieved referring to the inequality obtained in \cite{yuan2006model, simon2013sparse}. \\
\begin{equation}
	\parallel S \left( \beta_{jk}^0- t\bigtriangledown l(r_k^{(-j)}, \beta_{jk}^0), t\lambda_{1} {|\gamma_j^{*}| }^{\alpha} \right) \parallel_2 \leq t \cdot \lambda_{1g0}
\end{equation}
where $(S(z, \alpha \lambda))_j = sign(z_j) (|z_j| - \alpha \lambda)_{+}$.

Otherwise, by taking derivative of $\tilde{L_1(B)}$ with $\beta_{jk}$ we have 
\begin{equation}
	\left( 1+\frac{t \sum_{g \in G} \lambda_{1g}{\parallel \gamma_g^* \parallel_2^{\alpha} } } { \parallel \hat{B_g} \parallel_2 } \right) \hat{\beta_{jk}} = S \left( \beta_{jk}^0 - t \cdot \bigtriangledown l(r_k^{(-j)}, \beta_{jk}^0), t \lambda_{1}{|\gamma_j^{*}| }^{\alpha} \right)
\end{equation}

Then by taking the norm of both sides with respect to the group, we see that 
\begin{equation}
	\left( 1+\frac{t \sum_{G_{jk}} \lambda_{1g}{\parallel \gamma_g^* \parallel_2^{\alpha} } }{ \parallel \hat{B_g} \parallel_2 } \right) \parallel \hat{B_g} \parallel_2 = \parallel S ( \beta_{jk}^0 - t \cdot \bigtriangledown l(r_k^{(-j)}, \beta_{jk}^0), t \lambda_{1}{|\gamma_j^{*}| }^{\alpha} ) \parallel_2 
\end{equation}

Thus,
\begin{equation}
	\parallel \hat{B_g} \parallel_2 = \left( \parallel S(\beta_{jk}^0 - t \cdot \bigtriangledown l(r_k^{(-j)}, \beta_{jk}^0), t \lambda_{1}{|\gamma_j^{*}| }^{\alpha} ) \parallel_2 -t \sum_{G_{jk}} \lambda_{1g}{\parallel \gamma_g^* \parallel_2^{\alpha} } \right)_{+}
\end{equation}

Therefore we have 
\begin{equation}
	\hat{\beta_{jk}} = \left( 1 - \frac{t \sum_{G_{jk}} \lambda_{1g} {\parallel \gamma_g^* \parallel_2^{\alpha} } }{\parallel S(\beta_{jk}^0 - t \cdot \bigtriangledown l(r_k^{(-j)}, \beta_{jk}^0), t \lambda_{1}{|\gamma_j^{*}| }^{\alpha} ) \parallel_2}\right)_{+} S \left( \beta_{jk}^0 - t \cdot \bigtriangledown l(r_k^{(-j)}, \beta_{jk}^0), t \lambda_1{|\gamma_j^{*}| }^{\alpha} \right)
\end{equation}

Similarly, equations can be obtained for Model 2. 
\begin{equation}
	\hat{\gamma_{j}} = \left( 1 - \frac{t \sum_{G'_{j}} \lambda_{2g} {\parallel \beta_g^* \parallel_2^{\alpha} } }{\parallel S(\gamma_{j}^0 - t \cdot \bigtriangledown l(w^{(-j)}, \gamma_{j}^0), t \lambda_{2}{|\beta_{j \cdot}^{*}| }^{\alpha} ) \parallel_2}\right)_{+} S \left( \gamma_{j}^0 - t \cdot \bigtriangledown l(w^{(-j)}, \gamma_{j}^0), t \lambda_2{|\beta_{j \cdot}^{*}| }^{\alpha} \right)
\end{equation}
where $w^{(-j)} = Z - \sum_{j' \neq j} X_{j'}\gamma_{j'}$, as Z is a 1-dimensional vector, and so is the coefficient parameter $\gamma$.

If we iterate Equation (18), and recenter each pass at ${(\beta_{jk}^0)}_{new} = {(\hat{\beta_{jk}})}_{old} $, we will converge on the optimal solution for $\hat{\beta_{jk}}$ while fixing the values of other coefficients \cite{simon2013sparse, nesterov2013gradient}. Let $U(\beta_{jk}^0, t)$ denote the updated formula in  Equation (18), which represents the updated formula for each $\hat{\beta_{jk}}$ after obtaining $\beta_{jk}^0$ from the last step.
\begin{equation}
	U(\beta_{jk}^0, t) = \left( 1 - \frac{t \sum_{G_{jk}} \lambda_{1g} {\parallel \gamma_g^* \parallel_2^{\alpha} } }{\parallel S(\beta_{jk}^0 - t \cdot \bigtriangledown l(r_k^{(-j)}, \beta_{jk}^0), t \lambda_{1}{|\gamma_j^{*}| }^{\alpha} ) \parallel_2}\right)_{+} S \left( \beta_{jk}^0 - t \cdot \bigtriangledown l(r_k^{(-j)}, \beta_{jk}^0), t \lambda_1{|\gamma_j^{*}| }^{\alpha} \right)
\end{equation}

In the sparse group Lasso paper, the author proposed to fit using blockwise descent, which means they updated each group in a step while fixing the values of other groups. Besides, they employed an accelarated generalized gradient algorithm within each group \cite{simon2013sparse, simon2013blockwise}. The above equations we derive is to fit each coefficient of estimate at a time. Therefore, we apply the algorithm of coordinate descent to update each coefficient coordinate at a step while fixing all the other coefficients at their current values; and for each coefficient we employ an accelarated generalized gradient algorithm \cite{simon2013sparse, nesterov2013gradient, li2015multivariate}. As a combination, it's named as accelerated generalized coordinate descent algorithm. The convergence of the coordinate descent for multivariate group lasso and sparse group lasso have already been proven \cite{li2015multivariate, tseng2001convergence, wu2008coordinate, simon2013sparse}. 

The algorithm of fitting the joint model mainly consists of two parts, of which one part is to fit the single model, and the other part is to fit two models iteratively. The steps for fitting the single model includes an inner loop and an outer loop, where the inner loop means that we update each coefficient of estimate of the regression matrix one at a time, following equation 18 and 19 for each beta jk hat and gamma j hat correspondingly. After running the inner loop and obtain the estimate of the whole regression matrix for model 1 and model 2, we run the outer loop to check if any group satifies the inequality such as Inequality (14). If so, the group needs to be reassigned as zero. Then the final step is that we repeat the inner loop and outer loop to iterate until the coefficient of the estimated matrix reaches a pre-specified precision level. After fitting a single model, we could apply the coefficient estimate obtained to calculate the weight in the other model, and thus fitting two models iteratively to achieve the pre-specified precision level. The detailed steps are included in the Supplementary materials.

\subsection{Joint Model for Time-to-Event Outcome}
The main clinical data accessible to radiogenomics study is the information of patients' survival statuses \cite{bakr2018radiogenomic, aerts2014decoding, lee2008prediction, cancer2012comprehensive, collisson2014comprehensive}. A continuous clinical outcome linearly related to genomic data is rarely obtainable. Therefore, it is vital to build a survival model to enable associations between the clinical data with time-to-event outcome and genomic data. 

In Model 1, that is, imaging (Y) vs. gene expression (X) , the relationship remains to be the same as in Chapter 2. For Model 2, the relationship for survival ourcome (Z) vs. gene expression (X) is built based on the Cox model, and thus we maximize the logarithm of the partial likelihood \cite{simon2011regularization}, 
\begin{equation}
	l(\gamma) = \sum_{i \in D} [{x_{i}}^T \gamma - log(\sum_{j \in R_i} e^{x_{j}^T \gamma}  )]
\end{equation}

If we include the penalty term, then the optimization problem for Model 2 is defined as
\begin{equation}
	L_2(G) = \arg \min_B \frac{1}{n}[\sum_{i \in D} (log(\sum_{j \in R_i} e^{x_{j}^T \gamma}) - {x_{i}}^T \gamma )]  + \sum_{1 \leq j \leq p} \lambda_{2}|\beta_{j.}^{\ast}|^{\alpha} |\gamma_{j}| + 
	\sum_{g \in \textit{g}} \lambda_{2g}{\parallel \beta_g ^{\ast} \parallel}_2^{\alpha}  {\parallel G_{g} \parallel}_2
\end{equation}

We also employ the accelerated generalized gradient algorithm derived from Ref. \cite{simon2013sparse} for the continuous outcome. It has been proven that the algorithm can be extended to the model with convex loss functions \cite{simon2013sparse, nesterov2003introductory, nesterov2013gradient}, and it has already been shown that the partial likelihood of Cox model is convex \cite{wilson2021fenchel}. Therefore, referring to the extension in Ref. \cite{simon2013sparse}, we are also able to extend the algorithm to our Joint model with Cox regression.

The solutions for Model 1 is the same as Eqn (18), while for Model 2, equations are derived.
\begin{equation}
	\hat{\gamma_{j}} = \left( 1 - \frac{t \sum_{G'_{j}} \lambda_{2g} {\parallel \beta_g^* \parallel_2^{\alpha} } }{\parallel S(\gamma_{j}^0 - t \cdot \bigtriangledown l'(\gamma), t \lambda_{2}{|\beta_{j \cdot}^{*}| }^{\alpha} ) \parallel_2}\right)_{+} S \left( \gamma_{j}^0 - t \cdot \bigtriangledown l'(\gamma), t \lambda_2{|\beta_{j \cdot}^{*}| }^{\alpha} \right)
\end{equation}
where $l'(\gamma) = \sum_{i \in D} \left( log(\sum_{j \in R_i} e^{x_{j}^T \gamma}) - {x_{i}}^T \gamma \right)$. In Model 1, we defined $r_k^{(-j)}$ for convenience of symbol as it is a case of multivariate modeling. Here in Model 2, as $\gamma$ is a one-dimensional vector, we just take $l'(\gamma)$ as the unpenalized loss function and we can calculate $\bigtriangledown l'(\gamma)$, where $\bigtriangledown l'(\gamma) = \frac{1}{n} [\sum_{i \in D} (\frac{\sum_{j \in R_i} x_j^T e^{x_j^T \beta}}{\sum_{j \in R_i}e^{x_j^T \beta} } -x_i^T ) ]$.

Similarly, we have that the group containing $\gamma_{j}$ will be zero, i.e.$\hat{G_{g0}}=0$ if the following inequality is achieved referring to the inequality obtained in \cite{yuan2006model, simon2013sparse}. \\
\begin{equation}
	\parallel S \left( \gamma_{j}^0- t\bigtriangledown l'(\gamma), t\lambda_{2} {|\beta_{j \cdot}^{*}| }^{\alpha} \right) \parallel_2 \leq t \cdot \lambda_{g0}
\end{equation}

\section{Simulation Studies}
Simulation is very important to assess the performance of new models and the algorithms. In our simulation, we generate the dataset first so that we could compare the estimated results to the 'true' defined parameters. This is not accessible when applying the real data where we do not know the true parameters. Therefore, we perform simulation in order to ensure the correct coding, and test the performance of the new model and algorithm. 

\subsection{Continuous Outcome}
We perform comprehensive simulation studies to evaluate the performance of the joint model and compare the results with those obtained from some existing methods such as MSGLasso \cite{li2015multivariate}, SGL \cite{simon2013sparse}, and regular lasso \cite{tibshirani1996regression}. Since SGL and regular lasso only fits for the case of one-dimensional response, they are applied for Model 2 only. 

The covariates are generated from a multivariate normal distribution, i.e. $X_i^T$ $\sim$ $N_p(0, \Sigma_X)$, i = 1, ..., n, where the covariance matrix $\Sigma_X = diag(\Sigma_{g1}, ... , \Sigma_{g20})$ is block diagonal corresponds to each group of X, so that X features are correlated within groups, and independent between groups. For Model 1, the error term is $W=(w_1, ... , w_q) \in R^{n \times q}$, where each error term is generated from a normal distribution, that is, $w_k \sim N(0, \sigma_k^2 I_{n \times n})$, k =1, ..., q. B is the parameter for Model 1, where $B = (\beta_{jk})_{p\times q} \in R^{p \times q}$. In our simulation, we have p=200, and q=120, and define the index for grouping. We define twenty groups for gene expressions (X) and four groups for imaging (Y).  Then the responses Y are generated from the linear relationship Y=XB+W. For Model 2, the response variable is $Z \in R^{n}$, and thus $W'  \in R^n$ is the vector of error terms for Model 2, with $w' \sim N(0, \sigma'^2 I_{n \times n})$. G is the parameter for Model 2, where $G = (\gamma_{j})_{p\times 1} \in R^{p}$. Similarly, the response Z values are generated from the linear relationship Z=XG+W'. 

\begin{figure}[!h]
	\centering
	\includegraphics[scale=0.42]{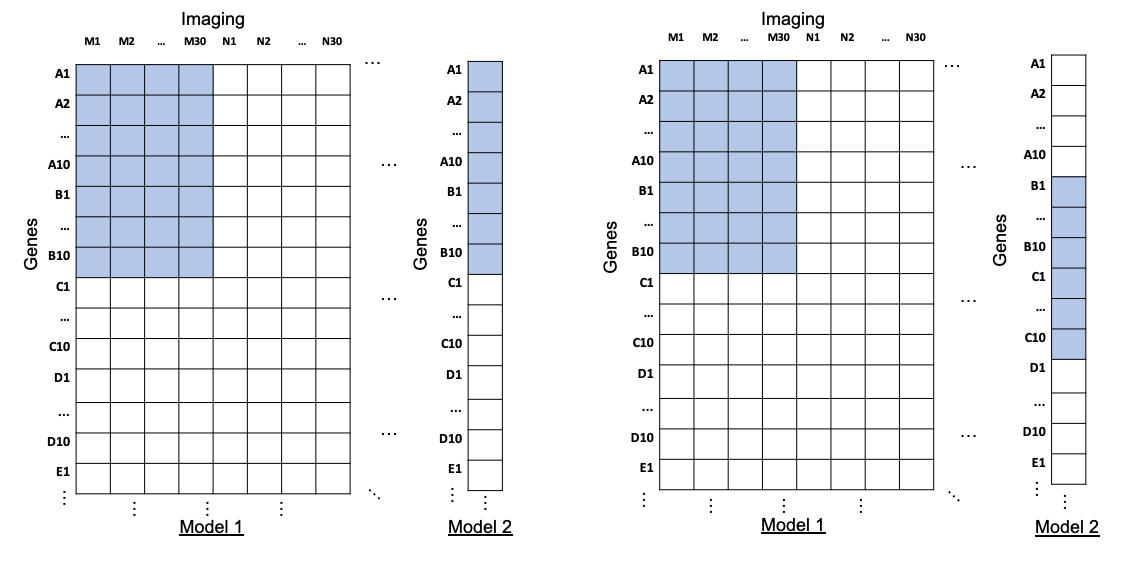}
	\caption{ Example of different overlapping cases for Model 1 and Model 2. 
		(a) Complete overlap (left figure); (b) Partial overlap (right figure)}
	\label{fig:P3}
\end{figure}

In our simulation, we set important features by defining coefficient values of B and G for Model 1 and Model 2, respectively. The important features will have non-zero values, while the values of unimportant features are zero. We considered four cases for each simulation setting, which is 100$\%$ overlap, 75$\%$ overlap, 50$\%$ overlap, and 25$\%$ overlap. The complete overlap is 100$\%$ overlap, which means that the important genes defined for Model 1 is exactly the same as in Model 2. The rest are the cases of partially overlap, that is, the important genes defined for Model 1 only partially overlap with those in Model 2, while 75$\%$, 50$\%$ and 25$\%$ define the percentage of overlapping of the important features. The cases of partial overlap are closer to the real case, since in real data, the genes associated with imaging features are most probably not exactly the same as the genes associated with the clinical outcome. Therefore, we test in both complete overlap and partial overlap cases to include effects from the overlapping extent. Figure \ref{fig:P3} illustrates an example of the complete and partial overlap cases. 

Besides, we consider four simulation scenarios, namely LS1, LS2, LS3 and LS4. The datasets generated in LS1 and LS2 have higher coefficient values than those in LS3 and LS4. Meanwhile, the sample size of the datasets in LS1 and LS3 is n=100, while n is 50 for LS2 and LS4. Hence, by comparing the results between LS1 vs LS3, and LS2 vs LS4, we can observe the effect of coefficient values. Furthermore, the effect of sample size can be observed from the comparison of the results between LS1 vs LS2, and LS3 vs LS4. Note that in each simulation setting, we include all four overlap cases from 100$\%$ overlap to 25$\%$ overlap.

To assess the prediction performance of our joint model, we calculate the prediction errors. In addition, since we have the information of the true parameter values in simulation, we can compare the obtained coefficient estimate with the true parameter values and calculate true positive rate (TPR) and true negative rate (TNR). We use relative reduction of prediction error (in short of RRPE) to estimate the prediction performance. Results of Model 2 are more important to us, because they represent the predictive performance results of the clinical outcome directly. Hence, we focus more on the results for Model 2 obtained from fitting the joint model.  For each simulation setting, we run fifty replications and calculate the average values of the following quantities to quantify the feature selections. \\
\begin{equation}
	\begin{aligned}
		TPR (Sensitivity) & = \frac{|{ \hat{\gamma_{j}} \neq 0 \textrm{ and } \gamma_{j} \neq 0 }|}{|{ \gamma_{j} \neq 0}|}, j=1,...,p \\
		TNR (Specificity) & = \frac{|{ \hat{\gamma_{j}} = 0 \textrm{ and } \gamma_{j} = 0 }|}{|{\gamma_{j} = 0}|}, j=1,...,p \\
		RRPE & = \frac{pe_{Null} - pe_{Method} }{pe_{Null}}
	\end{aligned}
\end{equation}
where $pe_{Null}$ is the prediction error of the null model, accounting for the variation of the dataset itself. $pe_{Method}$ refers to the prediction error with the $\hat{G}$ obtained from certain modeling method. Then, a higher $RRPE$ value indicates a better prediction performance, since it shows higher reduction of the prediction error via certain modeling method.

Besides the joint model, we also apply MSGLasso, SGL and regular Lasso methods for the purpose of comparison. Automatic parameter selection is implemented for each method. For the joint model, the optimal values of tuning parameters $\lambda_{1}$, $\lambda_{1g}$ for Model 1 and $\lambda_{2}$, $\lambda_{2g}$ for Model 2 are searched via 5-fold cross validation over a range of candidate values for each dataset. During the cross validation process, models are built using the training set with each set of $\lambda$ values from the candidate, and then residual sum of squares are calculated in the testing set. The set of $\lambda$ corresponding to the minimum residual sum of square from the fitting is then selected. For MSGLasso, SGL and regular Lasso method, we use 'MSGLasso.cv' function in 'MSGLasso' R package, 'cvSGL' function in 'SGL' R package, and 'cv.glmnet' function in 'glmnet' R package correspondingly, in order to find the optimal tuning parameters. 

\begin{figure}[!h]
	\centering
	\includegraphics[scale=0.34]{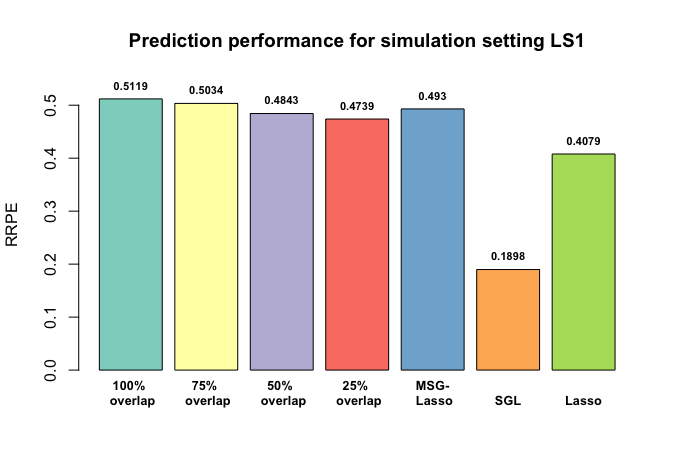}
	\includegraphics[scale=0.34]{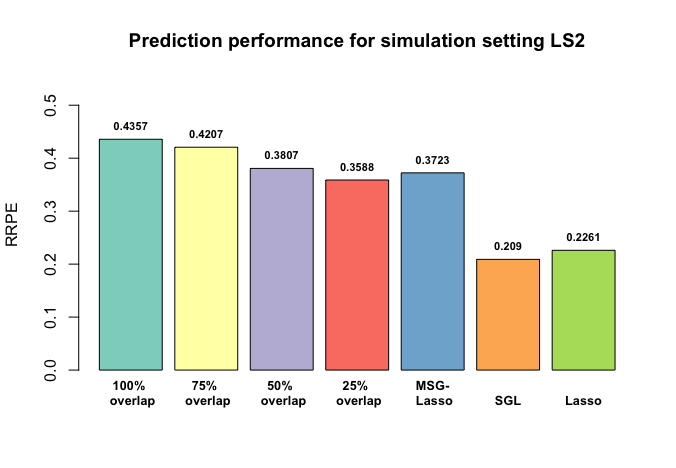}
	\includegraphics[scale=0.34]{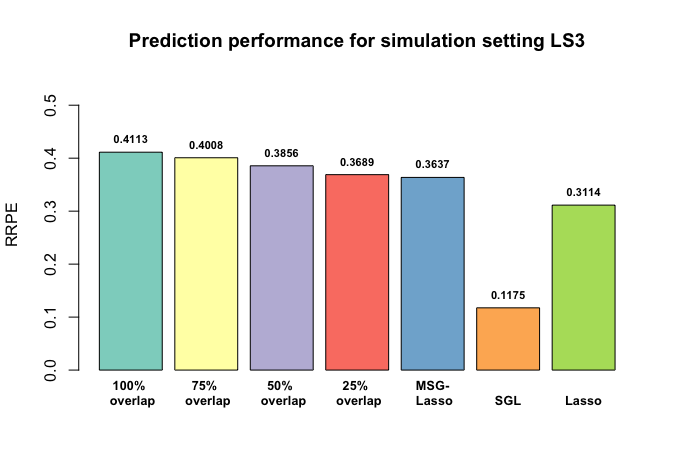}
	\includegraphics[scale=0.34]{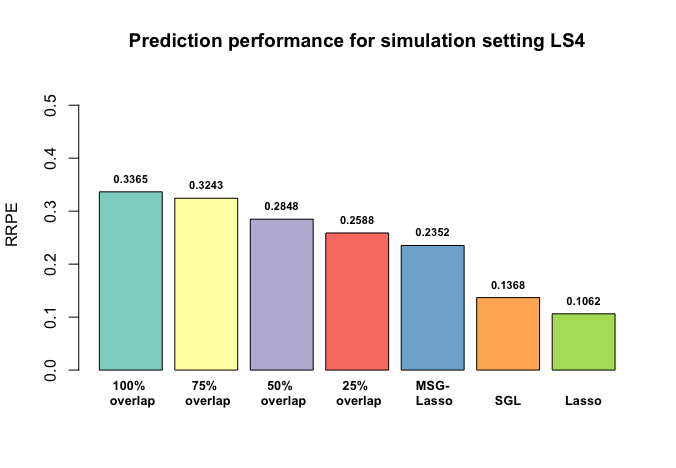}
\centering
	\caption{ Prediction performance for continuous outcome in each simulation setting. \\
		(a) Simulation setting LS1 (upper-left figure); (b) Simulation setting LS2 (upper-right figure); \\
		(c) Simulation setting LS3 (lower-left figure); (d) Simulation setting LS4 (lower-right figure)}
	\label{fig:P9}
\end{figure}

Figure \ref{fig:P9} summarizes the prediction performance of our joint model in all four overlapping cases corresponding to different simulation scenarios, as well as comparisons among the prediction performances obtained from the present method, MSGLasso, SGL, and regular lasso. 

It can be seen from Figure \ref{fig:P9} that the relative reduction of prediction error (RRPE) decreases with the decreasing of the overlapping extent. The cases of complete overlap yields the highest performance. As expected, a higher extent of overlap indicates more correct information conveyed from one model to the other. Additionally, in most cases even including those of partial overlap, the joint model performs better than the other three methods. MSGLasso method performs better than SGL and regular Lasso, but the joint model still outperforms MSGLasso in almost all scenarios and most partial overlapping cases. Especially when the sample size is small or the true coefficient value is low, the prediction performance of the joint model significantly outperforms that of MSGLasso.

Table \ref{tab:sim4} shows the results of feature selection for each simulation scenario, which indicates consistency with the RRPE results. With a less overlap extent, the performance of feature selection drops. The reason is that with decreasing the overlapping extent, the information conveyed between the two models will not always be precise and thus may lead to less selections of the true features or more selections of the false features. The trend is consistent with Figure \ref{fig:P9}. 
\begin{table}[!h]
	\caption{Prediction Performance: Feature Selection}
	\centering
	\begin{tabular}{llllllllllllllllllllllllllll}
		\hline
		& \multicolumn{2}{l}{XZ Model} \\ \hline
		Setting & TPR & TNR \\ \hline
		LS1-100$\%$ Overlap-Joint model &  0.9888 & 0.7892 \\
		LS1-75$\%$ Overlap-Joint model & 0.9798 & 0.7211 \\
		LS1-50$\%$ Overlap-Joint model &  0.9578 & 0.711 \\
		LS1-25$\%$ Overlap-Joint model &  0.9562 & 0.5969 \\
		LS1-MSGLasso  &   0.8446 & 0.9672 \\ 
		LS1-SGL  &   0.3374 & 0.9999 \\ 
		LS1-Lasso  &   0.544 & 0.9412 \\ \hline
		
		LS2-100$\%$ Overlap-Joint model &  0.9304 & 0.8546 \\
		LS2-75$\%$ Overlap-Joint model  &  0.9064 & 0.8243 \\
		LS2-50$\%$ Overlap-Joint model &  0.8778 & 0.7505 \\
		LS2-25$\%$ Overlap-Joint model &  0.8338 & 0.7232 \\
		LS2-MSGLasso  &   0.7724 & 0.9371 \\ 
		LS2-SGL  &   0.2422 & 0.9984 \\ 
		LS2-Lasso  &   0.5312 & 0.9315 \\ \hline
		
		LS3-100$\%$ Overlap-Joint model &  0.9826 & 0.7443 \\
		LS3-75$\%$ Overlap-Joint model & 0.9762 & 0.6707 \\
		LS3-50$\%$ Overlap-Joint model &  0.9556 & 0.6304 \\
		LS3-25$\%$ Overlap-Joint model &  0.949 & 0.5262 \\
		LS3-MSGLasso  &   0.6902 & 0.9515 \\ 
		LS3-SGL  &   0.304 & 0.9997 \\ 
		LS3-Lasso  &   0.3738 & 0.9396 \\  \hline  
		
		LS4-100$\%$ Overlap-Joint model  &  0.9344 & 0.8147 \\
		LS4-75$\%$ Overlap-Joint model &  0.909 & 0.7854 \\
		LS4-50$\%$ Overlap-Joint model &  0.8824 & 0.7135 \\
		LS4-25$\%$ Overlap-Joint model &  0.8378 & 0.6567 \\
		LS4-MSGLasso  &   0.6354 & 0.9378 \\ 
		LS4-SGL  &   0.2208 & 0.9969 \\ 
		LS4-Lasso  &   0.3186 & 0.9432 \\ \hline
	\end{tabular}
	\label{tab:sim4}
\end{table}

Besides, from the comparison in the TPR and TNR values of the different modeling methods in Table \ref{tab:sim4}, we observe that TPRs of the joint model are higher than those of all other modeling methods regardless of the overlap extent, while TNRs are lower than those of other modeling methods. This indicates that the joint model tends to select more features. The main reason is that we add a weight term such as ${|\gamma_j^{*}| }^{\alpha}$ obtained from the other model to the penalty term $\lambda_1$, which leads to decreased penalty values. Therefore, more features will be selected. In addition, the joint model enables us to capture information conveyed between the two models, which will help to avoid missing findings of the significant features. It can be seen from the prediction performance results (Figure \ref{fig:P9}) that capturing more features is beneficial for the enhancement of prediction performance, since RRPE of the joint model is higher.

Furthermore, it can be observed from the results across all simulation scenarios that the joint model performs better than the other methods regardless of the sample size or effect size. In LS1, the joint model with 100$\%$ and 75$\%$ overlap performs better than MSGLasso. The difference in performance enlarges with reducing the effect size or sample size. For example, in LS4 where true coefficient values and sample size are the smallest, the joint model outperforms other methods significantly. From LS1 vs LS2, and LS3 vs LS4, we can see that lower sample size yields lower RRPE and TPR for all modeling methods. Particularly, MSGLasso, and regular Lasso are sensitive to the sample size, as RRPE drops significantly with the lower sample size. From LS1 vs LS3, and LS2 vs LS4, we observe that decreasing the true coefficient value while constructing the dataset leads to a reduction of both RRPE and TPR. The trend is consistent with our expectation, since the coefficient value indicates whether the signal is strong enough. Overall, it has been demonstrated an improvement of the prediction performance by applying the joint model within each numerical setting of the simulation studies.

\subsection{Time-to-Event Outcome}
To evaluate whether our joint model can outperform the reported models that fits a single Cox model, such as Cox model using SGL package \cite{simon2013sparse}, lasso Cox model with glmnet package \cite{simon2011regularization}, and regular Cox model \cite{survival-package, survival-book}, we conducted comprehensive simulation studies with data generated with survival outcome. 

For Model 2, the data frame is built for survival analysis. We generate survival time T from the exponential distribution by defining the hazard as rate in the exponential distribution, where $h_i(t) = h_0(t) \cdot e^{x_i^T \gamma}$ for each observation i = 1,...,n. To mimic a real-world scenario, we obtain our baseline hazard based on a real radiogenomic dataset from Ref. \cite{bakr2018radiogenomic}. We generate the censoring time C and generally fix the consoring rate at 20$\%$. Then we define $Z = min(T, C)$ and the censoring indicator $\delta$ as $\delta = 1$ if $T \leq C$ and 0 otherwise. 

Estimation of the simulation is done to check whether the desired prediction performance as well as the precision of the feature selection is reached. We referred to the reported survival-ROC method to estimate time-dependent prediction performance for the survival outcome. By applying the package 'survivalROC', we can obtain AUC value at different time points of interest \cite{heagerty2005survival, heagerty2000time, heagerty2013package}. The following quantities are calculated by taking an average of fifty replications for each case in each simulation setting.
\begin{equation}
	\begin{aligned}
		\textnormal{Prediction performance of Model 2}, auc_2(t) \\
		TPR (Sensitivity) & = \frac{|{ \hat{\gamma_{j}} \neq 0 \textrm{ and } \gamma_{j} \neq 0 }|}{|{ \gamma_{j} \neq 0}|}, j=1,...,p \\
		TNR (Specificity) & = \frac{|{ \hat{\gamma_{j}} = 0 \textrm{ and } \gamma_{j} = 0 }|}{|{\gamma_{j} = 0}|}, j=1,...,p
	\end{aligned}
\end{equation}

We consider four scenarios, where the scenario 1 and scenario 2 have higher coefficients than scenario 3 and scenario 4, and meanwhile, scenario 1 and 3 are with the sample size n=100 while scenario 2 and 4 are with the sample size n=50. Combinations of the sample sizes and coefficient values are summarized in Table \ref{tab:sim2}. 

\begin{table}[h]
	\centering
	\caption{Scenarios settings}
	
	\begin{tabular}{lllll}
		\hline
		Setting & Sample size (n) & Coefficient values  \\ \hline 
		S1 & 100 & 0.3  \\ 
		S2 &  50 & 0.3 \\ 
		S3 &  100 & 0.25 \\ 
		S4 & 50 & 0.25 \\ \hline
	\end{tabular}
	\label{tab:sim2}
\end{table}

\begin{figure}[!h]
	\centering
	\includegraphics[scale=0.35]{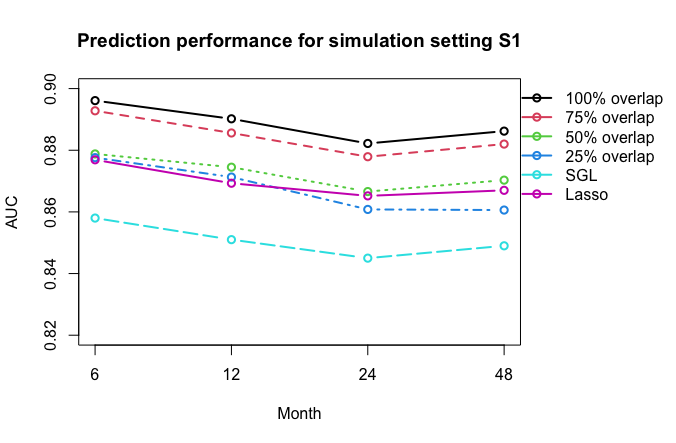}
	\includegraphics[scale=0.35]{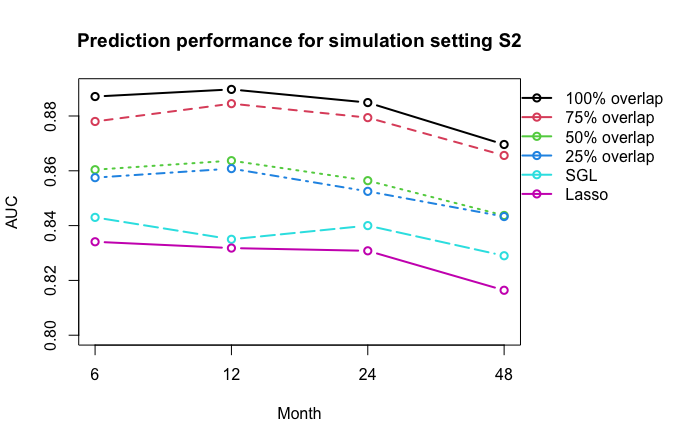}
	\includegraphics[scale=0.35]{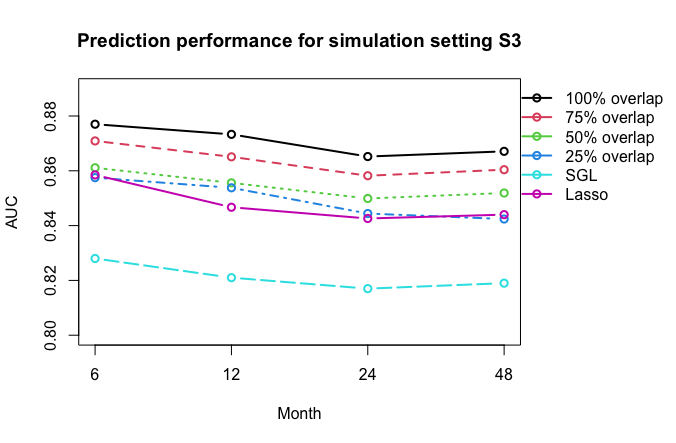}
	\includegraphics[scale=0.35]{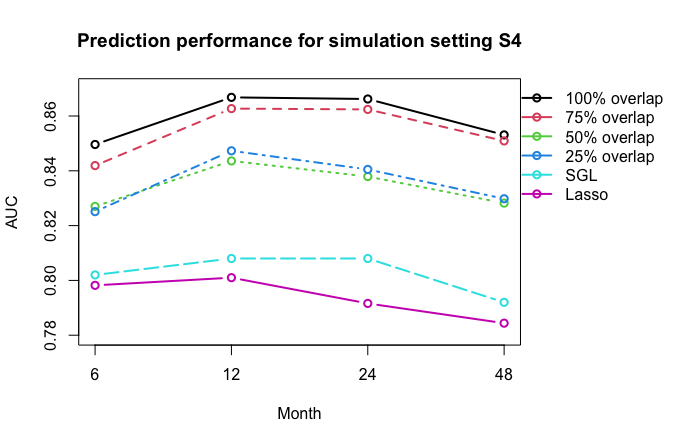}
	\caption{ Prediction performance for the survival outcome in each simulation setting. \\
		(a) Simulation setting S1 (upper-left figure); (b) Simulation setting S2 (upper-right figure); \\
		(c) Simulation setting S3 (lower-left figure); (d) Simulation setting S4 (lower-right figure)}
	\label{fig:P4}
\end{figure}

As a result, Figure \ref{fig:P4} shows the prediction performance at different time points of interest for each simulation setting. Time points of interest are chosen based on the survival time generated and are also typical time points of interest for this kind of study. 

From Figure \ref{fig:P4}, we can see that AUC decreases with decreasing the overlapping extent. AUC is the highest for the complete overlapping case as expected, since the information conveyed from one model should be completely useful for the other model. Obviously, under the case of partial overlap, the information conveyed between the two models could be misleading, since not all the significant features overlap. Even though the information only partially overlaps, our joint model still shows better prediction performance than SGL model for Cox regression with the light blue dotted line and Lasso model for Cox regression in almost all simulation settings. 

Besides, results of the feature selection for Model 2 is shown in Table \ref{tab:sim3}. It can be seen that the feature selection results are mostly consistent with the prediction performance. In each simulation setting, feature selection for complete overlap outperforms those of other cases. For the cases of partial overlap, true positive rate (TPR) decreases when the degree of overlapping decreases. TPR represents the ratio of important features that have been correctly selected. As for TNR representing the ratio of correctly identified non-important features, overall it decreases from complete overlap (100$\%$ overlap) to 25$\%$ overlap, but for 75$\%$ and 50$\%$ overlap the trend is not clear. This is resulted from the difference in findings of the optimal tuning parameters $\lambda_{1}$, $\lambda_{1g}$, $\lambda_{2}$, and $\lambda_{2g}$. Smaller optimal tuning parameters are chosen for the case of 75$\%$ overlap, which means smaller penalty value for all parameters. As a result, more features will be chosen, and thus TNR will be lower. Combining Table \ref{tab:sim3} with the prediction performance results shown in Figure \ref{fig:P4}, we note that capturing more important features in the joint model is beneficial for improving the prediction performance, as AUC of the joint model is higher than those of other models.

In addition, Table \ref{tab:sim3} shows that TPR of our joint model is consistently higher than that of SGL model and Lasso model for Cox regression in each simulation setting, which means that the joint model improves the feature selection of the important features. According to the prediction performance characterized with AUC as shown in Figure \ref{fig:P4}, we can see that the improvement in feature selection helps enhance the prediction performance as well. Though TNR of the joint model is not always higher than those of SGL and Lasso model for Cox regression, it is still reasonable because we add another term in addition to the original tuning parameters $\lambda_{2}$ and $\lambda_{2g}$, in order to convey information from the other model. Some information will be incorrect especially for the case of partial overlap. As a result, joint model may select the features that should not be important, thus TNR is relatively lower.

\begin{table}[!h]
	\caption{Prediction Performance: Feature Selection}
	\centering
	\begin{tabular}{llllllllllllllllllllllllllll}
		\hline
		& \multicolumn{2}{l}{XZ Model} \\ \hline
		Setting & TPR & TNR \\ \hline
		S1-100$\%$ Overlap-Joint model &  0.9938 & 0.9374 \\
		S1-75$\%$ Overlap-Joint model & 0.9716 & 0.7751 \\
		S1-50$\%$ Overlap-Joint model &  0.9452 & 0.8924 \\
		S1-25$\%$ Overlap-Joint model &  0.8924 & 0.7203 \\
		S1-SGL   &   0.7876 & 0.7304 \\ 
		S1-Lasso   &   0.7392 & 0.9648 \\ \hline
		
		S2-100$\%$ Overlap-Joint model &  0.9798 & 0.9297 \\
		S2-75$\%$ Overlap-Joint model  &  0.9434 & 0.7123 \\
		S2-50$\%$ Overlap-Joint model &  0.873 & 0.8257 \\
		S2-25$\%$ Overlap-Joint model &  0.8666 & 0.6697 \\
		S2-SGL &  0.6143 & 0.8292  \\ 
		S2-Lasso  &  0.4498 & 0.9748  \\ \hline
		
		S3-100$\%$ Overlap-Joint model &  0.986 & 0.9264 \\
		S3-75$\%$ Overlap-Joint model & 0.9676 & 0.7576 \\
		S3-50$\%$ Overlap-Joint model &  0.9426 & 0.8726 \\
		S3-25$\%$ Overlap-Joint model &  0.8894 & 0.6888 \\
		S3-SGL &  0.7380 & 0.7144  \\ 
		S3-Lasso &  0.6702 & 0.9669  \\  \hline  
		
		S4-100$\%$ Overlap-Joint model  &  0.9634 & 0.923 \\
		S4-75$\%$ Overlap-Joint model &  0.931 & 0.7054 \\
		S4-50$\%$ Overlap-Joint model &  0.845 & 0.8549 \\
		S4-25$\%$ Overlap-Joint model &  0.8532 & 0.6974 \\
		S4-SGL &  0.5627 & 0.8159  \\ 
		S4-Lasso &  0.3856 & 0.9762  \\ \hline
	\end{tabular}
	\label{tab:sim3}
\end{table}

Both AUC and feature selection results show that the joint model performs better than the other methods in each simulation scenario. In S1, the joint model with 100$\%$, 75$\%$ and 50$\%$ overlap have higher AUC than Lasso model, which mostly coincide with the 25$\%$ overlap curve. Across the different scenarios, we can observe that the difference in performance enlarges with reducing the effect size or sample size. In specific, in S2 and S4 the AUCs of the joint model are much higher than those of other methods. Furthermore, it can be seen from both AUC and feature selection results that the performance of Lasso model is influenced largely by the sample size. From simulation setting S1 to S2, and from S3 to S4, the performance has become inferior as the sample size is cut by half.  The simulation study has demonstrated a significant improvement of the prediction performance by applying joint model within each simulation setting.

\section{Real Data Analysis}
To further demonstrate the applicability of the method we proposed, we evaluate the joint model using real-world data analysis and compare the results with those obtained from other existing methods. We conduct studies on two datasets of non-small cell lung cancer (NSCLC). One dataset is a published dataset with gene expression data, imaging data, and survival outcome data \cite{bakr2018radiogenomic}. The other dataset is combined from two parts, of which one part is from  data generated by the TCGA Research Network: \url{https://www.cancer.gov/tcga}, and the rest also from the published paper (Ref. \cite{bakr2018radiogenomic}). 

The first dataset Bakr et al. used is from an published article which provides a complete radiogenomic dataset from a Non-Small Cell Lung Cancer (NSCLC) cohort \cite{bakr2018radiogenomic}. The dataset comprises Computed Tomography (CT), Positron Emission Tomography (PET)/CT images, segmentation maps of tumors in the CT scans, gene expression microarrays and RNA sequencing data from samples of surgically excised tumor tissue, as well as the clinical data, including survival outcomes \cite{bakr2018radiogenomic}. For the CT images with outlined region of interests (ROIs) already, a quantitative imaging feature pipeline is applied to generate 107 computational imaging features corresponding to 135 patients. The image features are extracted using an open source python library named PyRadiomics \cite{van2017computational}. The resultant 107 features belongs to First Order Statistics, Shape, Gray Level Cooccurrence Matrix, Gray Level Run Length Matrix, Gray Level Size Zone Matrix, Neighboring Gray Tone Difference Matrix and Gray Level Dependence Matrix feature classes. Radiomics features are extracted from tumor ROIs.

As for the gene expression dataset, originally it is of the unit fragments per kilobase of exon per million mapped fragments (FPKM). We convert it to the unit of TPM to adjust for the gene length \cite{conesa2016survey}. Furthermore, a standardization has been performed for both gene expression dataset and imaging dataset, because Lasso regression puts constraints on the size of the coefficients associated to each variable with a penalty term, and the value will be influenced by the magnitude of each variable. It is therefore necessary to center and/or standardize the variables \cite{tibshirani1996regression, tibshirani2012strong}. Clinical outcome includes overall survival and censoring status. Intercept of the three parts leads to a sample size of 109 patients of this common cohort. 

We notice that the combined dataset is of different histologic subtype. We divide the dataset according to Ref. \cite{langer2010evolving, selvaggi2009histologic} into the histology of Adenocarcinoma (LUAD) and Squamous cell carcinoma (LUSC) which are available in our dataset. As a result, there are 82 patients with the histology of Adenocarcinoma, while there are only 24 patients with the histology of Squamous cell carcinoma. Considering the need of adequate sample size, we apply our joint model on the dataset with LUAD only for further analysis.  

Grouping can be performed for this dataset. Gene expression features can be grouped by pathways. Here we refer to the KEGG pathway for NSCLC  \cite{kanehisa2017kegg}.  There are 7 pathways, such as Ras signaling pathway, ErbB signaling pathway, calcium signaling pathway, cell cycle, etc. Besides, imaging features are grouped by shape, size, and other parameters, and there are also 7 imaging groups. 

\begin{figure}[!h]
	\centering
	\includegraphics[scale=0.5]{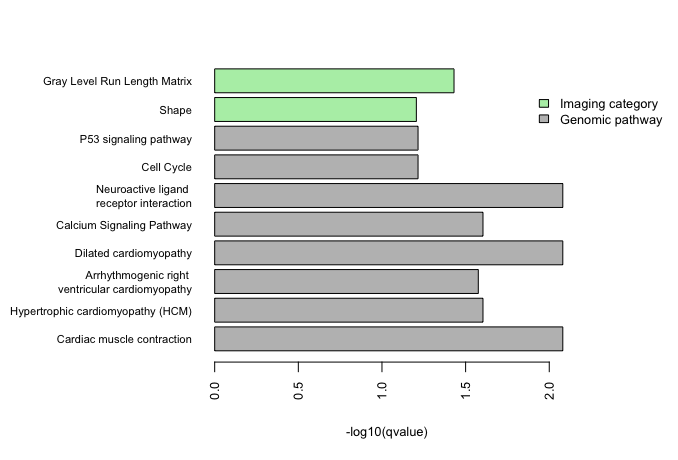}
	\caption{ Associations between imaging/genomic features and clinical outcome. }
	\label{fig:P7}
\end{figure}

We perform some exploratory analysis on this dataset. A gene set enrichment analysis (GSEA) has been implemented to evaluate the association between overall survival and a category of imaging features, or transcriptional activity of a KEGG pathway. The association is shown in Figure \ref{fig:P7}. Imaging features or gene expressions are ranked based on the model coefficient of the univariate Cox regression model. The false discovery rate q-value is obtained based on the Benjamini-Hochberg method.  

\begin{figure}[!h]
	\centering
	\includegraphics[scale=0.75]{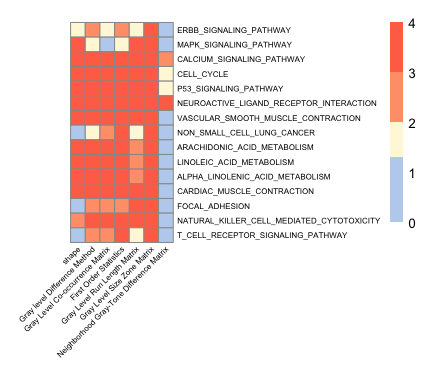}
	\caption{ Associations between imaging and genomic features.}
	\label{fig:P8}
\end{figure}

From Figure \ref{fig:P7} we could see that there exist some genomic pathways and categories of imaging features that have statistically significant associations with the clinical outcome. In addition, we also confirm that associations between imaging and genomic features do exist. We then perform a gene set enrichment analysis (GSEA) to assess the association between an imaging feature and a KEGG pathway, and Figure \ref{fig:P8} displays some of the associations between imaging groups and genomic pathways.  We rank the genes based on the Pearson's correlation coefficient between gene expression and imaging feature. The Fisher's method is used to obtain a combined p-value for each pathway and category. We also take a negative logarithm of the q-values and thus the higher value indicates closer associations. The figure shows that there are indeed some statistically significant associations between imaging and genomic features. Therefore, it supports the idea of our joint model that three parts of the radiogenomics data are associated. We will use gene expression data to predict patients' survival outcome, and including the imaging data may contribute to better selection of informative features and enhancement of prediction. 

The second dataset is a combination from TCGA-LUAD and the dataset Bakr et al, which will be thereafter called TCGA-LUAD for convenience. Since the sample size of the dataset that satisfies the condition of both containing three parts and being with the same cohort is limited, we consider combining dataset from different cohorts. Gene expression data and clinical outcome of patients with NSCLC of Adenocarcinoma (LUAD) are from the open source TCGA research network. Thus, we can fit Model 2 of the joint model with solely this part, while fit Model 1 with imaging vs. gene expression from the dataset Bakr et al., as long as we match gene expression features from the two parts. 

The sample size of TCGA-LUAD is over five hundred. Similar processing has been done on Gene expression data (referred as X2 for TCGA-LUAD) and Survival outcome (Z) from TCGA. The original unit of gene expression dataset is of fragments per kilobase of exon per million mapped fragments (FPKM). We convert it to the unit of TPM in adjustment for the gene length, and then implement standardization to reduce the size effect. 

\subsection{Analysis on the dataset Bakr et al.}
We fit the dataset Bakr et al.with our joint model using the above-mentioned algorithm. In comparison, we also fit with Lasso Cox using the 'glmnet' package in R \cite{simon2011regularization}. We note that MSGLasso model \cite{li2015multivariate} cannot be applied, since it does not work for the time-to-event outcome. Additionally, though SGL method can be applied for the time-to-event data, it does not work here for the dataset with overlapped grouping. As introduced before, gene expression features can be grouped by pathway, where different pathways can possibly contain the same gene features. 

\begin{figure}[!h]
	\centering
	\includegraphics[scale=0.4]{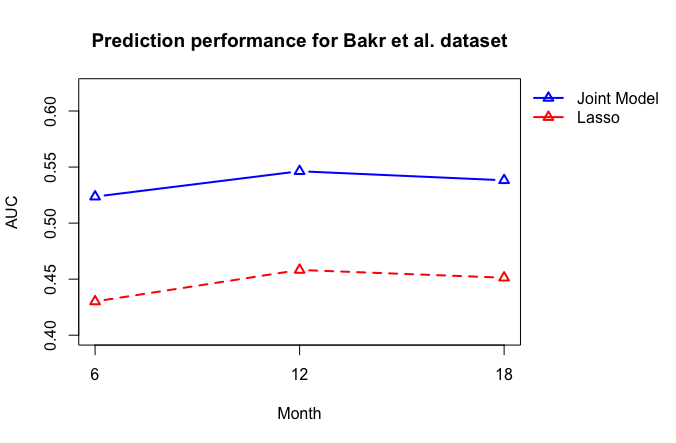}
	\caption{ Prediction performance for the survival outcome of the dataset Bakr et al.}
	\label{fig:P5}
\end{figure}

As a result, Figure \ref{fig:P5} displays the prediction performance characterized by AUC at 6 months, 12 months and 18 months. AUCs are calculated by the 'survivalROC' package mentioned before in Section 3.1. Overall, AUC of the joint model is higher than the Lasso method, indicating a better prediction performance. However, AUCs are only over 0.50, and are not satisfactory. The relatively low AUC values are mainly due to the small sample size. We initially used the whole dataset to fit, but results show that combinations of the histology did not work well. According to the exploratory analysis, where we apply the univariate Cox model to study the associations between the survival outcome and each gene expression feature, as well as associations between outcome and imaging feature, rare significant associations were found, as shown in Appendix C. Thus, we divide the dataset into histology of Adenocarcinoma  and Squamous cell carcinoma \cite{langer2010evolving, selvaggi2009histologic}, and there are only 82 patients which is comparably inadequate on contrary to the number of features.

\subsection{Analysis on the dataset TCGA-LUAD}
TCGA-LUAD provides a dataset with much larger sample size. Therefore, we could fit Model 2 of the joint model using the TCGA dataset. As for fitting Model 1, we can still apply the gene expression and imaging dataset from the dataset Bakr et al. \cite{bakr2018radiogenomic}. The dataset and the corresponding model applied are summarized in Table \ref{tab:real2}. Note that we need to match features of X1 and X2  in order to convey the information from one model to the other via the coefficient estimates and thus the joint model will be applicable. 

\begin{table}[!h]
	\caption{TCGA-LUAD}
	\centering
	\begin{tabular}{llllllllllllllllllllllllllll}
		\hline
		Model & Dataset \\ \hline
		Model 1 & \shortstack{Gene expression data (X1) and Imaging data (Y) from the dataset Bakr et al. \cite{bakr2018radiogenomic}}    \\
		Model 2 & \shortstack{Gene expression data (X2) and Survival outcome (Z) from TCGA-LUAD} \\ \hline
	\end{tabular}
	\label{tab:real2}
\end{table}

\begin{figure}[!h]
	\centering
	\includegraphics[scale=0.4]{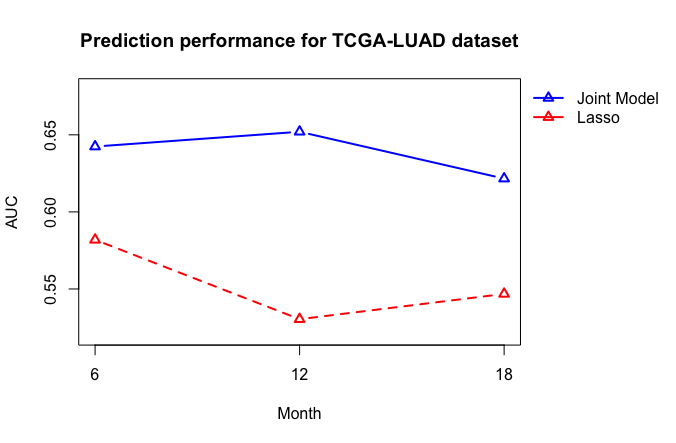}
	\caption{ Prediction performance for the survival outcome of TCGA-LUAD}
	\label{fig:P6}
\end{figure}

We fit this dataset with the joint model as well as Lasso Cox model, and similarly, we use the survival-ROC method to calculate AUC at several time points of interest. Figure \ref{fig:P6} shows the prediction performance. A big improvement is shown here, as all the AUC values of the joint model are over 0.6, with an average value of around 0.65, demonstrating a satisfactory prediction performance. It is also shown that the performance of the joint model significantly surpasses that of Lasso model for Cox regression. The results not only show an improvement in predicting the survival outcome realized by using the joint model, but also shows the feasibility of radiogenomics modeling on the dataset from different sources, which expands the application of the joint model. 

In a word, applying the joint model on the TCGA-LUAD dataset improves the prediction performance. From the prediction results we can see that AUC decreases if we predict later time points of interest. AUC of the joint model at 12 months reaches over 0.65, and drops to 0.62 at 18 months. The reason could be that we used the baseline gene expression data to predict the survival outcome, leading to a reduction in prediction precision for later time points. 

Our joint model allows the use of separate datasets to fit the two models, which assists in resolving the existing limitations of datasets availability. As there are many public datasets only has genomic information but not imaging information, our approach enables researchers to adopt such datasets in combination with other datasets wherever imaging data is available. Therefore, the sample size of analysis dataset can be enlarged, and the performance of predicting clinical outcome can be improved.

\section{Discussion}
In summary, our joint model integrates both genomic and imaging to predict the clinical outcome. We also extend the current group lasso model and algorithm to the multivariate response and overlapped grouping for survival models, and therefore solve the limitations of current existing methods. Importantly, in both the simulation studies and real data analysis, we can see that the joint model demonstrates superior prediction performances when compared with existing methods, and we also see an improvement in feature selection. In addition, the joint model allows the use of separate datasets to fit the two models. There are many public datasets only has genomic information but not imaging information, and our joint model enables researchers to use such datasets combining with other datasets where the imaging data is available. This can help solve the current problem of limited available dataset.

Though the joint model leads to a better prediction performance from the real data analysis, the corresponding AUC is not very satisfactory, with values around 0.65 for the dataset TCGA-LUAD. One possible reason is that there could be more genomic pathways related. For example, from Figures \ref{fig:P7} and \ref{fig:P8}, we can see some other KEGG pathways that have statistically significant association with the survival outcome or imaging features. Inevitably, including more pathways will lead to including more genes in the dataset, and as a result, computational efficiency might be influenced. We will consider including more pathways in the next step, and meanwhile try to balance the computing capacity of the joint model. 

Besides, some covariates may influence the survival outcome, such as age, gender, smoking and cancer stage. Such covariates can be added to the joint model, which may help improve the prediction performance. We will continue work on the joint model to allow the input of covariates. Penalty terms will not be added for the covariate variables, as covariates terms should be excluded from feature selection while fitting the joint model.  

Since it takes some time to build the two models iteratively, especially with a large number of features, thus an alternative way is to fit imaging vs genomics separately, and then take the selected important features from the fit as a prior information while fitting survival outcome vs. genomics.  However, the precision of feature selection might be reduced, since it will be a one-time modeling instead of the iterative modeling.

Furthermore, an possible alternative for the joint model could be using imaging features to predict the clinical outcome in Model 2 of the joint model. Model 1 for imaging feature vs. geneomics could remain the same, and from the fitting, we could select imaging features instead, based on the coefficient of estimate obtained. Then fit clinical outcome with imaging feature in Model 2. 

\section*{Acknowledgements}
This work was supported by National Institutes of Health [5P20GM103436-15] and the Biostatistics and Bioinformatics Shared Resource Facilities of the
University of Kentucky Markey Cancer Center [P30CA177558].
	

\newpage

	\section*{Supplement to \newline
	'Multivariate Sparse Group Lasso Joint Model for Radiogenomics Data'}
Tiantian Zeng \textsuperscript{1*}, 
Md Selim \textsuperscript{2},
Jie Zhang \textsuperscript{3},
Arnold Stromberg \textsuperscript{1},
Jin Chen\textsuperscript{2,4}, 
Chi Wang \textsuperscript{1,4*}

\subsection*{Proof of Theorem 2.1}
We denote our unpenalized loss function by
\begin{equation*}
	\begin{split}
		l(B) & = \frac{1}{2} \parallel Y-XB \parallel^2_2  \\
		& = \frac{1}{2} \sum_{i=1}^{n} \sum_{k=1}^{q}(y_{ik}-\sum_{j=1}^{p}x_{ij}\beta_{jk})^2	
	\end{split}
\end{equation*}
Then 
\begin{equation*}
	\frac{\partial l(B) }{\partial \beta_{jk}} = -x_j^T (Y-XB)_{.k}
\end{equation*}
Define  
\begin{equation}\tag{A1}
	S_{jk} = x_j^T (Y-XB^{(-j)})_{.k}
\end{equation}
then $ -S_{jk} = -x_j^T (Y-XB^{(-j)})_{.k}$, \\
then $ \frac{\partial l(B) }{\partial \beta_{jk}} - (-S_{jk}) = x_j^T [(XB)_{.k} - (XB^{(-j)})_{.k} ] = \parallel x_j \parallel_2^2 \beta_{jk} $. \\
Thus we have 
\begin{equation*}
	\frac{\partial l(B) }{\partial \beta_{jk}} = -S_{jk} +  \parallel x_j \parallel_2^2 \beta_{jk}
\end{equation*}

Then Equation (3) in Section 2.1 becomes
\begin{equation*}
	L_1(B) = \frac{1}{n} l(B) + \sum_{1 \leq j \leq p, 1 \leq k \leq q}\lambda_1{|\gamma_j^{*}| }^{\alpha} |\beta_{jk}| + 
	\sum_{g \in \textit{g}} \lambda_{1g}{\parallel \gamma_g^* \parallel_2^{\alpha} } {\parallel B_{g} \parallel}_2
\end{equation*}

We derived solutions to the above equation of $L_1(B)$ based on the Karush-Kuhn-Tucker conditions \cite{yuan2006model, hastie2015statistical}. 

For any $\beta_{jk}$ where $L_1(B)$ is differentiable at $\beta_{jk}$, we have
\begin{equation*}
	\frac{\partial L_1(B)}{\partial \beta_{jk}} =  -\frac{1}{n} S_{jk} + \frac{1}{n} \parallel x_j \parallel_2^2 \beta_{jk} + \lambda_1{|\gamma_j^{*}| }^{\alpha} sgn(\beta_{jk}) + \sum_{g_{jk}} \lambda_{1g}\parallel \gamma_g^* \parallel_2^{\alpha} \beta_{jk}/{\parallel B_{g} \parallel}_2
\end{equation*}
where $g_{jk}$ is the group that $\beta_{jk}$ belongs to. \\
If $\beta_{jk} >0$, then $ 	\frac{\partial L_1(B)}{\partial \beta_{jk}} =  -\frac{1}{n} S_{jk} + \frac{1}{n} \parallel x_j \parallel_2^2 \beta_{jk} + \lambda_1{|\gamma_j^{*}| }^{\alpha} + \sum_{g_{jk}} \lambda_{1g}\parallel \gamma_g^* \parallel_2^{\alpha} \beta_{jk}/{\parallel B_{g} \parallel}_2 $. \\
In this case $ \frac{\partial L_1(B)}{\partial \beta_{jk}}  \geq 0$ if and only if 
\begin{equation*}
	\beta_{jk} \geq \frac{S_{jk} -n \lambda_1 {|\gamma_j^{*}| }^{\alpha} } { \parallel x_j \parallel_2^2 + n \sum_{g_{jk}} \lambda_{1g}\parallel \gamma_g^* \parallel_2^{\alpha} /{\parallel B_{g} \parallel}_2 } \coloneqq \hat{\beta_{jk}^+}
\end{equation*}
and also, $ \frac{\partial L_1(B)}{\partial \beta_{jk}}  < 0$ if and only if 
\begin{equation*}
	\beta_{jk} < \hat{\beta_{jk}^+}
\end{equation*}
That is, if $\beta_{jk} >0$, $ L_1(B)$ is monotone increasing when $\beta_{jk} > \hat{\beta_{jk}^+}$, and monotone decreasing when $\beta_{jk} < \hat{\beta_{jk}^+}$.  Therefore $ \hat{\beta_{jk}^+} $ minimizes $ L_1(B)$  wrt $\beta_{jk}$ when $\beta_{jk} > 0$, i.e. 
\begin{equation} \tag{A2}
	\hat{\beta_{jk}^+} = 
	\begin{cases}
		\frac{S_{jk} -n \lambda_1 {|\gamma_j^{*}| }^{\alpha} } { \parallel x_j \parallel_2^2 + n \sum_{g_{jk}} \lambda_{1g}\parallel \gamma_g^* \parallel_2^{\alpha} /{\parallel \hat{B}_{g} \parallel}_2 } , & \text{if}\ S_{jk} >n \lambda_1 {|\gamma_j^{*}| }^{\alpha}  \\
		0, & \text{otherwise}
	\end{cases}
\end{equation}
Similarly, we can obtain $	\hat{\beta_{jk}^-} $ minimizes $ L_1(B)$  wrt $\beta_{jk}$ when $\beta_{jk} < 0$,
\begin{equation}\tag{A3}
	\hat{\beta_{jk}^-} = 
	\begin{cases}
		\frac{S_{jk} +n \lambda_1 {|\gamma_j^{*}| }^{\alpha} } { \parallel x_j \parallel_2^2 + n \sum_{g_{jk}} \lambda_{1g}\parallel \gamma_g^* \parallel_2^{\alpha} /{\parallel \hat{B}_{g} \parallel}_2 } , & \text{if}\ S_{jk}  < -n \lambda_1 {|\gamma_j^{*}| }^{\alpha}  \\
		0, & \text{otherwise}
	\end{cases}
\end{equation}
Combining Eqn (A2) and (A3), we have when $\beta_{jk} \neq 0$,
\begin{equation} \tag{A4}
	\hat{\beta_{jk}} = \frac{ sgn(S_{jk}) (S_{jk} - n \lambda_1 {|\gamma_j^{*}| }^{\alpha})_+ } { \parallel x_j \parallel_2^2 + n \sum_{g_{jk}} \lambda_{1g}\parallel \gamma_g^* \parallel_2^{\alpha} /{\parallel \hat{B}_{g} \parallel}_2 }
\end{equation}
Eqn (A4) is Eqn (5) in Section 2.1. Similarly, for Model 2, we could obtain Eqn (7). 

When $\beta_{jk} = 0$, there are two possible cases depending on whether all the groups which contain $\beta_{jk}$ are zero groups or not. If none of the groups containing $\beta_{jk}$ is a zero group, then
\begin{equation*}
	\frac{\partial L_1(B)}{\partial \beta_{jk}} =  -\frac{1}{n} S_{jk} + \frac{1}{n} \parallel x_j \parallel_2^2 \beta_{jk} + \lambda_1{|\gamma_j^{*}| }^{\alpha} u + \sum_{g_{jk}} \lambda_{1g}\parallel \gamma_g^* \parallel_2^{\alpha} \beta_{jk}/{\parallel B_{g} \parallel}_2
\end{equation*}
where $|u| \le 1$.  Then a similar discussion to the case based on whether $u > 0$ or $u \leq 0$ yields the same expression (A4). 

Denote $ \textit{g}_{jk}  = \{g: \beta_{jk} \in B_g \in \textit{g} \}$, that is, the groups that $\beta_{jk}$ belongs to. Recall that in Section 2.1, we define the group structure as $ \textit{g} = \{ B_1, ..., B_G \} $. 

Secondly, if some of the groups contain $\beta_{jk}$ are zero groups, then neither $| \beta_{jk} |$ nor $ \parallel B_g \parallel_2 $ is differentiable at zero. Let $\textit{g}_{jk}^{\otimes} = \{g: \beta_{jk} \in B_g \in \textit{g}, where \parallel B_g \parallel_2 > 0 \}$. Then for any zero groups $B_{g0}$ in $ \textit{g}_{jk} $, and for all $\beta_{jk} \in B_{g0}$, $\hat{\beta_{jk}}$ needs to satisfy the subgradient equation: 
\begin{equation*}
	\frac{1}{n} S_{jk} - \frac{1}{n} \parallel x_j \parallel_2^2 \hat{\beta_{jk} } = \lambda_1{|\gamma_j^{*}| }^{\alpha} u + \lambda_{g0} v_{jk} + \sum_{ \textit{g}_{jk}^{\otimes} } \lambda_{1g}\parallel \gamma_g^* \parallel_2^{\alpha} \hat{\beta_{jk} }/{\parallel B_{g} \parallel}_2
\end{equation*}
where u is the subgradient scalar for the $l_1$ norm $| \beta_{jk} |$, and v is the subgradient vector for the $l_2$ norm $ \parallel B_g \parallel_2 $, with constraints $|u|\leq 1 $ and $ \parallel v \parallel_2 \leq 1 $, and $\lambda_{g0}$ is the tuning parameter associated with $B_{g0}$. 

By combining the subgradient equation with basic algebra, we get that $\hat{B_{g0}}=0$ if
\begin{equation*}
	\sqrt{\sum\limits_{{jk:\beta_{jk} \in B_{g0}}} (|S_{jk}|/n - \lambda_{1}|\gamma_j ^{\ast}|^{\alpha} )_{+}^2 } \leq \lambda_{g0}
\end{equation*}  

For Model 2, the steps are exactly the same as for Model 1, and similarly we obtain the Inequality (8) to check for the groups.

\subsection*{Detailed Steps of Accelarated Generalized Coordinate Descent Algorithm }
The steps for updating the coefficient estimate as following. \\
Step 1. (Inner loop) \\
(a) Start with $\beta^{(jk, l)} = \theta^{(jk, l)} = \beta_{jk}^0 $, step size t=1, and counter l=1;  \\
(b) Update gradient g by $g = \bigtriangledown l(r_k^{(-j)}, \beta^{(jk, l)})$ ; \\
(c) Optimize step size by iterating t=0.8*t until 
\begin{equation*}
	l(U(\beta^{(jk, l)}, t)) \leq l(\beta^{(jk, l)}) + g^T \bigtriangleup_{(l,t)} + \frac{1}{2t} \parallel \bigtriangleup_{(l,t)} \parallel_2^2
\end{equation*}
Here we search for the step size t in order to hold the descent direction of the majorization scheme. \\
(d) update $\theta^{jk, l}$ by 
\begin{equation*}
	\theta^{(jk, l+1)} \leftarrow U(\beta^{(jk, l)}, t)
\end{equation*}
Note that $U(\beta^{(jk, l)}, t)$ follows Equation (20). \\
(e) update the center via a Nesterov step. This Nesterov-style momentum updates will enable us to leverage some higher order information while only calculate gradients. (derived from Simon's group lasso paper and Nesterov step). 
\begin{equation*}
	\beta^{(jk,l+1)} \leftarrow \theta^{(jk,l)} + \frac{l}{l+3} (\theta^{(jk, l+1)} - \theta^{(jk,l)})
\end{equation*}
5) set l=l+1 \\
where $\bigtriangleup_{(l,t)}$ is the change between our old solution and new solution, $\bigtriangleup_{(l,t)} = U(\beta^{(jk,l)}, t) - \beta^{(jk,l)} $ \\

Step 2. (Outer loop) If any of the group $B_g \in \textit{g}$ containing $\beta_{jk}$ satisfies the following, then the entire group is estimated at zero.   
\begin{equation*}
	\parallel S \left( \beta_{jk}^0- t\bigtriangledown l(r_k^{(-j)}, \beta_{jk}^0), t\lambda_{1} {|\gamma_j^{*}| }^{\alpha} \right) \parallel_2 \leq t \cdot \lambda_{1g0}
\end{equation*}

Step 3.  Repeat step 1 and step 2 for all $j \in {1,...,p}$ and $k \in {1,...,q}$, and iterate until $\parallel \hat{B}^{(m)} - \hat{B}^{(m-1)} \parallel$ reaches a pre-specified precision level, where m refers to the index count of each iteration. 

For Model 2, the estimation of coefficients are also updated via Step 1 to Step 3, following Equation (19), and thus we can obtain $\hat{\gamma_{j}}$ for each j=1,...,p.

The above algorithm demonstrates the steps of obtaining the coefficient estimate for Model 1 and Model 2 correspondingly. The joint model conveys the coefficient obtained from one model to the other. The steps for the joint model is as follows. \\ 
Initially, set $\gamma_j^{*}=1$ for all j=1,...p and $\parallel \gamma_g^* \parallel_2 = 1$ for and group g. \\
Step A. Follow step 1 to step 3 above to fit Model 1, and obtain the estimate of coefficient matrix for Model 1, i.e. $\hat{B}$. \\
Step B. Calculate $\beta_{j \cdot} ^*$ for each j=1,...,p, by $\beta_{j.}^* = -\tilde{\beta}_{j \cdot} / (\frac{1}{P}\sum_{j=1}^P (-\tilde{\beta}_{j \cdot}))$, while $\tilde{\beta}_{j \cdot} = \log_{10}  (\max \limits _ {1 \leq k \leq q} |\beta_{jk}|)$. Meanwhile, define $\tilde{\beta}_{j \cdot}=-0.01$ if $\tilde{\beta}_{j \cdot}>=0$, and $\tilde{\beta}_{j \cdot} = -2$ if $\tilde{\beta}_{j \cdot}<-2$ in order to reduce the difference of numerical values, since it works as a penalty term in the other model that will influence the feature selection. At the same time, calculate the group term $\parallel \beta_g^* \parallel_2$, where $\parallel \beta_g^* \parallel_2 =  {\parallel \tilde{\beta_g}  \parallel}_2 / (\frac{1}{G}\sum_{g=1}^G {\parallel \tilde{\beta_g}  \parallel}_2)$ and $\parallel \tilde{\beta}_g \parallel_2 = \sqrt{\sum_{\tilde{\beta_{j \cdot}}  \in B_g}  \tilde{\beta_{j \cdot}}^2}$. \\
Step C. Apply the values of parameter from Step B to fit Model 2, also by following step 1 to step 3 above. $\hat{G}$ can be obtained for Model 2. \\
Step D. Calculate $\gamma_j^{*}$ for all j=1,...,p, by $\gamma_j^{*}= -\tilde{\gamma}_{j} / (\frac{1}{P}\sum_{j=1}^P (-\tilde{\gamma}_{j}))$, while $\tilde{\gamma}_{j} = \log_{10} |\gamma_{j}|$. Meanwhile, define $\tilde{\gamma}_{j}=-0.01$ if $\tilde{\gamma}_{j} \geq 0$, and $\tilde{\gamma}_{j} = -2$ if $\tilde{\gamma}_{j}<-2$, so as to reduce the gap of numerical values. At the same time, calculate the group term $\parallel \gamma_g^* \parallel_2$, by $\parallel \gamma_g^* \parallel_2 = {\parallel \tilde{\gamma_g}  \parallel}_2 / (\frac{1}{G}\sum_{g=1}^G {\parallel \tilde{\gamma_g}  \parallel}_2)$ and ${\parallel \tilde{\gamma_g}  \parallel}_2 = \sqrt{\sum_{\gamma_{j}  \in G_g} \tilde{\gamma_j}^{2}}$. \\
Step E. Iterate repeatedly via Step A to Step D above to achieve the pre-specified precision level for both models. 

For a given dataset with the three parts (genomic, imaging and outcome), we will be able to fit the model jointly following the above steps, and estimate accordingly the coefficients for each model.

\end{document}